\documentclass[reprint,aps,prl,groupedaddress,amsmath,amssymb]{revtex4-2}
\usepackage{graphicx,color,upgreek}
\usepackage{mathtools}
\usepackage{amsmath}
\usepackage{mathtools}
\usepackage{layouts}
\usepackage{float}
\usepackage{hyperref}
\hypersetup{
    colorlinks=true,
    linkcolor=blue,
    filecolor=magenta,      
    urlcolor=cyan,
    pdftitle={Overleaf Example},
    pdfpagemode=FullScreen,
    }
\usepackage{makecell}
\usepackage[abs]{overpic}
\usepackage{helvet}
\usepackage{graphicx,color,natbib}
\usepackage{placeins}
\usepackage[dvipsnames]{xcolor}
\usepackage{array, booktabs, makecell}
\usepackage{siunitx}
\usepackage{longtable}
\usepackage{tabularx}
\usepackage{booktabs}
\let\oldAA\AA
\renewcommand{\AA}{\text{\normalfont\oldAA}}
\usepackage{textcomp}
\usepackage{color}
\usepackage{amssymb,amsmath} 
\usepackage{siunitx}

\usepackage{algpseudocode}
\usepackage{algorithmicx}
\usepackage{algorithm}
\usepackage{lmodern}
\usepackage{amsmath, xparse}
\usepackage{colortbl}
\usepackage{soul}
\usepackage{xr}
\usepackage{xspace}

\definecolor{cream}{RGB}{222,217,201}
\definecolor{myblue}{rgb}{0.003921,0.07058,0.47450}
\definecolor{teal}{rgb}{0.0,0.5664,0.5742}
\definecolor{strawberry}{rgb}{1.0,0.0,0.5}
\definecolor{darkstrawberry}{rgb}{0.875,0.0,0.4375}
\definecolor{royallBlue}{rgb}{0.0039,0.0705,0.4745}
\definecolor{blueberry}{rgb}{0.015686275,0.2,1}
\definecolor{dukeblue}{RGB}{1,33,105}
\definecolor{darkgreen}{RGB}{0,150,0}
\definecolor{magenta}{RGB}{255,0,255}
\definecolor{red}{RGB}{255,0,0}

\newcommand{\numcolloidschallenging}{4\xspace}
\newcommand{\unet}{U-net\xspace}
\newcommand{\attunet}{Attention U-net\xspace}

\begin{document}

\title{Colloidoscope: Detecting Dense Colloids in 3d with Deep Learning}

\author{Abdelwahab Kawafi}
\affiliation{School of Physiology, Pharmacology, and Neuroscience, University of Bristol, Bristol, BS8 1TD, United Kingdom}

\author{Lars K\"{u}rten}
\affiliation{Gulliver UMR CNRS 7083, ESPCI Paris, Universit\'{e} PSL, 75005 Paris, France}

\author{Levke Ortlieb}
\affiliation{H. H. Wills Physics Laboratory, University of Bristol, Bristol BS8 1TL, United Kingdom.}

\author{Yushi Yang}
\affiliation{H. H. Wills Physics Laboratory, University of Bristol, Bristol BS8 1TL, United Kingdom.}

\author{Abraham Mauleon Amieva}
\affiliation{H. H. Wills Physics Laboratory, University of Bristol, Bristol BS8 1TL, United Kingdom.}

\author{James E. Hallett}
\affiliation{Department of Chemistry, School of Chemistry, Food and Pharmacy, University of Reading, Reading RG6 6AD, United Kingdom.}

\author{C.Patrick Royall}
\affiliation{Gulliver UMR CNRS 7083, ESPCI Paris, Universit\'{e} PSL, 75005 Paris, France}
\affiliation{H. H. Wills Physics Laboratory, University of Bristol, Bristol BS8 1TL, United Kingdom.}

\footnotetext{\textit{$^{a}$~School of Physiology, Pharmacology, and Neuroscience, University of Bristol, Bristol, BS8 1TD, United Kingdom.}}
\footnotetext{\textit{$^{b}$~Gulliver UMR CNRS 7083, ESPCI Paris, Universit\'{e} PSL, 75005 Paris, France.}}

\begin{abstract}
Colloidoscope is a deep learning pipeline employing a 3D residual \unet architecture, designed to enhance the tracking of dense colloidal suspensions through confocal microscopy. This methodology uses a simulated training dataset that reflects a wide array of real-world imaging conditions, specifically targeting high colloid volume fraction and low-contrast scenarios where traditional detection methods struggle. Central to our approach is the use of experimental signal-to-noise ratio (SNR), contrast-to-noise ratio (CNR), and point-spread-functions (PSFs) to accurately quantify and simulate the experimental data. Our findings reveal that Colloidoscope achieves superior recall in particle detection (finds more particles) compared to conventional heuristic methods. Simultaneously, high precision is maintained (high fraction of true positives.) The model demonstrates a notable robustness to photobleached samples, thereby prolonging the imaging time and number of frames than may be acquired. Furthermore, Colloidoscope maintains small scale resolution sufficient to classify local structural motifs. Evaluated across both simulated and experimental datasets, Colloidoscope brings the advancements in computer vision offered by deep learning to particle tracking at high volume fractions. We offer a promising tool for researchers in the soft matter community, this model is deployed and available to use pretrained \url{https://github.com/wahabk/colloidoscope.}
\end{abstract}

\maketitle

\section{Introduction}
\label{sectionIntroduction}

In addition to being important materials in their own right~\cite{russel}, colloids follow the same laws of statistical mechanics as do atoms and molecules~\cite{evans2019}, yet with a timescale amenable to experiments and a lengthscale appropriate for analysis with optical techniques. Of the latter, light scattering of colloids then provides comparable information to x-ray scattering of atoms and molecules~\cite{pusey2002ch1}. However \emph{real space analysis} using optical -- and particularly 3D confocal -- microscopy is challenging. Tracking the colloidal particles in space and time, so-called \emph{particle-resolved studies} (PRS), provide insight into phenomena generic to condensed matter, yet are hard to access with scattering methods~\cite{ivlev,gasser2009,hunter2012,yunker2014,royall2023}. For this reason, particle--resolved studies of colloidal systems have long held a fascination well 
beyond the direct interest in colloids as soft materials. Among the highlights are the first direct observation of the birth of a crystal nucleus~\cite{gasser2001} and dynamical heterogeneity~\cite{weeks2000,kegel2000} understood to be important in the long standing challenge of the glass transition. The advent of active colloidal systems lends further insight to be gained from particle--resolved studies~\cite{bechinger2016,mallory2016,palacci2010,bricard2013,mauleonamieva2023sciadv,buttinoni2013,sakai2020}.

\subsection{Classical colloid detection}
\label{sectionClassical}

\begin{figure*}[!ht]
\centering
\includegraphics[width=0.8\textwidth]{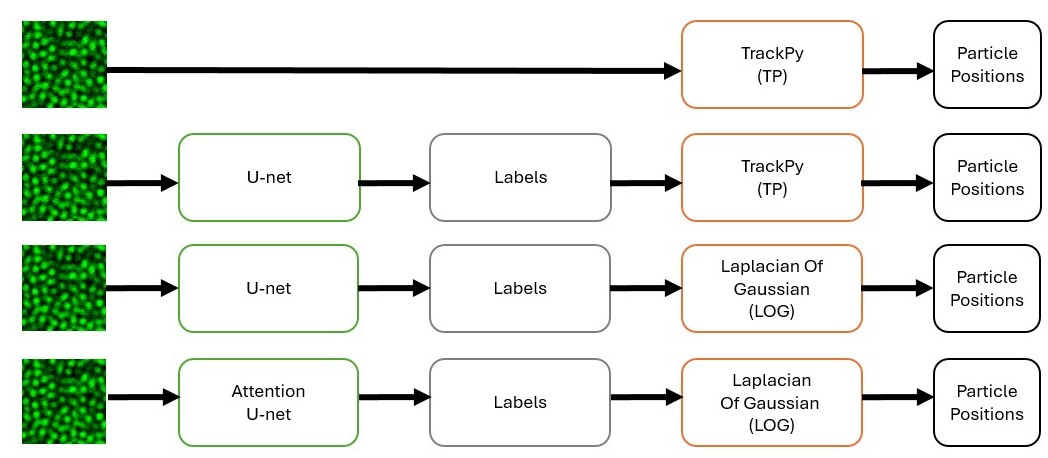}
\caption[Schematic]{Colloidoscope Schematic: Trackpy (TP) works end to end, extracting the particle positions from the image. Colloidoscope uses a \unet 
to label the image as a gaussian heatmap, then TP or LOG can be used as a postprocessing method to extract the particle positions. An \attunet is also tested. }
\label{figSchematic}
\end{figure*}

At a technical level, the seminal work of Crocker and Grier~\cite{crocker1996} for imaging of (quasi) 2D colloidal model systems has formed the backbone of much colloid tracking since. Indeed it is fair to say that the challenges of tracking colloids in 2D have been largely addressed, at least in comparison to the those presented by tracking colloidal particles in 3D ~\cite{ivlev}. However as successful as particle--resolved studies of colloids undoubtedly are~\cite{ivlev,gasser2009,hunter2012,yunker2014,royall2023}, major issues remain. For example, not all particles are tracked (low recall), ``ghost particles'' are erroneously identified (low precision), and all coordinates which are detected are subject to a tracking error (low Average Precision (AP)). These issues all become rather more significant when smaller particles or suspensions at high volume fraction are analysed~\cite{royall2023,hallett2018}. Recently, the case for using smaller colloids has been made~\cite{dauchot2022}, which provides access to new phenomena, particularly in the context of the glass transition~\cite{hallett2018,hallett2020,ortlieb2023}. The Crocker and Grier method has been implemented in the popular Python package Trackpy~\cite{allan2023}. 
From here on this will be referred to as TP.

Crocker and Grier ~\cite{crocker1996} is a simple algorithm that takes the ``pseudo diameter'' (known as $w$) to be larger than the true radius but smaller than the true diameter. This is used to preprocess the image with a boxcar filter followed by grayscale dilation ~\cite{jain1989} to detect the centroids. Then the integrated sphere brightness and the distance between the geometric centre and the centre of brightness mass of the particle is used to remove ``ghost particles'' or False Positives.

More sophisticated approaches are also available ~\cite{bierbaum2017, leocmach2013}. Among the pioneers of the field,~\citet{vanblaaderen1995} sought to improve the tracking accuracy in the axial direction by fitting a Gaussian to the integrated intensities in each plane that constituted a given colloid. This early work examined hard-sphere glasses and crystals, i.e., colloidal solids in which diffusion could be neglected ~\cite{vanblaaderen1995}. Even though this method made tracking easier, the slow scan rates needed to acquire images with low noise left dynamical information beyond reach. More recently,~\citet{jenkins2011} pushed the limits of the technique to identify contacts between colloids through ultra-high precision coordinate location. Such precision was achieved by first determining an empirical image of a colloid, which could then be compared to the original image. Further systematic improvements were made by~\citet{gao2009}. Despite advances in imaging technologies such as STimulated Emisison via Depletion (STED) ~\cite{hellnobel,hell2007}, the software to detect the colloids has lagged behind.

Tracking colloids is often specimen dependant~\cite{dong2022,midtvedt2022}, leading to fragmented tracking methods used by different researchers. A further issue regarding particle tracking, that perhaps has received less attention than it might, is that it can be something of a ``dark art''. Analysis can be slow, tedious, expensive, and remains subjective due to the extensive user-tuned parameters in tracking software. There exists a need for the implementation of state of the art computer vision methodologies for confocal microscopy that can be as specimen agnostic as possible. Many datapoints are discarded because they are of insufficient quality for tracking methodologies. Better detection algorithms can contribute to a lower laser light, line averaging, resolution, and increasing framerates. And in this way, we show that detecting smaller particles without needing improved imaging hardware becomes possible.

\subsection{Deep Learning for Dense Colloidal Suspensions}
\label{sectionDeepLearning}

Before proceeding, we define two key characteristics of how successful the particle tracking is.
\begin{itemize}
    \item \emph{Recall} is the proportion of particles in the sample which are successfully detected.
    \item \emph{Precision} is the fraction of those detected particles which correspond to (real) particles in the sample.
\end{itemize}

As proposed by ~\citet{bailey2022}, one might imagine that machine-learning techniques could be applied to the vexing problem of obtaining particle coordinates from microscopy images. It is natural to enquire as to the opportunities offered by the explosion in machine learning (ML). 
ML has revolutionised image analysis in many settings from earth observation ~\cite{yuan2021} to molecular imaging~\cite{ge2020}. However, one area in which there is potential for development is in tracking of concentrated colloidal dispersions imaged in 3D.

Deep learning for computer vision classicaly involves convolutional neural networks (CNNs). Convolutions are used to replace the fully connected layers of the network by constraining neurons to a specific kernel size - e.g. 3x3. This induces a ``receptive field'' where a hierarchy of neurons analyse each section of an image separately. This reduces the total number of connections and, therefore, parameters in the model, as well as provides translational invariance. We do not consider vision transformers due to their data inefficiency ~\cite{dosovitskiy2021}.

A common method of using CNNs is in semantic segmentation. This is where every pixel of the image is labelled with a specific class, e.g. particle (foreground) or background. Bounding box detection and instance segmentation ~\cite{redmon2016,he2017} - where each instance of a label is detected separately, are popular in object detection. However, these methods are not amenable to dense or 3D detection ~\cite{kar2022}. This is usually attributed to region proposal methods requiring fully connected layers that lose the benefit of translational invariance that fully convolutional models possess. Most particle tracking methods use one of two approaches. Firstly semantic segmentation can be used, a common method for cell counting ~\cite{midtvedt2020}. This can be combined with a post-processing method such as a watershed which finds distinct regions to extract the particle positions.
Secondly, a heatmap regression approach can be used. This is the approach used for colloidoscope and we note that this is a popular approach for \emph{dense} pose-prediction ~\cite{cao2016}.

\subsection{Previous Deep Learning Approaches}
\label{sectionGroundTruth}

A deep learning method for PRS has been developed by ~\citet{newby2018}. This consists of a small fully convolutional and recurrent network trained on simulated training data. This method outperformed all other methods in high signal to noise ratios and at high recall. Moreover it benefits from a recurrent approach where all frames are detected by the same model. Newby \emph{et al.}'s method also managed to track bacteria, hinting at the possibility to generalise this method to heterogenous datasets. However, this lacked the ability to predict the positions of particles in suspensions at high volume fraction. Other methods exist for the detection of colloids using deep learning, but focus on other microscopy approaches such as holographic imaging or STochastic Optical Reconstruction Microscopy (STORM)~\cite{nehme2018,altman2020}.

This challenge of applying ML methods to PRS has thus received some attention. DeepTrack~\cite{midtvedt2020} has made great headway in tracking a wider variety of colloids with different shapes such as crescents. However this method is focused towards dilute samples in 2D. As alluded to above, for colloidal systems, 2D analysis in particular is quite well dealt with using conventional methods. We believe that the case of 3D, hampered by the relatively poor resolution in the axial direction of confocal microscopy, is a more serious challenge, especially for concentrated systems. Furthermore, DeepTrack does not provide pretrained hyperparameters and takes the approach of training a unique model for each specimen. When compared to ~\citet{newby2018}, this brings up the question of: \emph{how much} to genaralise? Generalising to different datasets can aid in overfitting, but will it sacrifice precision or recall?

There is still not a sufficient approach neural network that accurately locates individual particles in a \emph{dense} colloidal suspension in 3D, to the best of our knowledge. This is because obtaining labeled data, where the 3D fluorescent images are annotated with the exact 3D locations for each individual colloid, is difficult. The 3D locations from both existing tracking software and human operators are not accurate enough to train a sufficient model.

\subsection{Ground Truth Deficiency}
\label{sectionGroundTruth}

From an ML perspective, particle resolved studies (PRS) is a \emph{ground truth deficient} problem, by which it is meant that there does not exist training data in which the solution (the coordinates of the particles) is precisely known from experimental data. To address the lack of ground truth for applications such as particle tracking, it is possible to use self-supervised learning, for instance~\citet{midtvedt2022} use geometric symmetry in sphere detection along with self-distillation to learn from each prediction. Another approach is to use generative adverserial neural-networks (GANs) to create the training dataset~\cite{zargari2024}.

The data deficiency can also be addressed by using simulated data to train the neural network. Applying molecular dynamics (MD) computer simulation, we can generate an ensemble of coordinates for densely packed particles, by repeatedly collecting particle locations from a simulated trajectory in the phase space. With the simulated coordinates, we can further simulate their corresponding microscopy images, by convolving the pulse functions located at the coordinates with suitable kernels such as the Point Spread Function (PSF)~\cite{bierbaum2017}. A simulation approach facilitates investigating the model's sensitivity to the training data parameters.

\subsection{Colloidoscope}
\label{sectionColloidoscope}

Here we present a deep learning model to detect colloidal particles. We provide comparison with current methods of particle tracking from the wide application of confocal microscopy. We aim to: 
(\emph{i}) provide a means to track 
colloids at high volume fraction ($\phi$) with both high precision and recall.
(\emph{ii}) simulate a training dataset of spherical colloids of a wide variety but a focus on glasses.
(\emph{iii}) train a model on the simulated dataset, for a tracking by detection method.
In this way, the \unet massively enhances the quality of the image from a point of view of particle tracking. The resulting \emph{labelled image} enables particle tracking to be more effective than the raw (or deconvolved) image.
Figure \ref{figSchematic} shows the proposed benchmark and pipelines investigated. We begin by training and testing on the simulation, fine tuning the model on 4 challenging images of colloids. These colloids are at different sizes and volume fractions, including a deconvoluted image. Finally we will test on a relatively simpler image to find how a \unet can improve on TP.

\section{Methodology}
\label{sectionMethodology}

This work is available at colloidoscope github.com/wahabk/colloidoscope.

\subsection{Model training, hyperparameter tuning, and data augmentation}
\label{sectionTraining}

The model for this project is a simple 3D \unet~\cite{cicek2016}. A \unet comprises a fully convolutional autoencoder with long skip connections from the encoder to decoder that utilizes dense features without losing spatial information. A residual encoder is used~\cite{drozdzal2016} with a block approach to test model depth. Loss functions are crucial for training CNNs. They are particularly crucial for small and or dense particle detection, both of which are true for this application. Since this model takes a heatmap approach, a simple $L_1$ loss function is used.

The volume at which a sample is imaged forms a compromise between resolution, working distance, as well as photobleaching. A confocal image is typically 16 megapixels (256 pixels cubed) or larger which constitutes our input. Such memory demand exceeds the capacities for most personal computing platforms. Tiling -- or a patch based pipeline -- is crucial for this work, where each image is broken into smaller regions of interest for predictions. Tiling inference saves on GPU memory and allows for a larger extent of the image in the $z$ direction (i.e. a greater number of $xy$ image planes) when compared to a 2D model with slice-wise inferencing. This method is also more amenable to a larger batch size during training (for improved normalisation), as well as test-time augmentation.

\attunet~\cite{oktay2018} was tested to measure its performance against the standard \unet. The original work introduces attention as a tool to combat class imbalance and organ localisation. Attention could still be crucial for dense detection, and would be dependant on the labelling method (binary segmentation vs. heatmap). For instance if each particle prediction is a Gaussian with a small fixed width (e.g. 5 pixels), class imbalance could still be an issue where most of the volume is background when $\phi$ is  low. Conversely in high volume fractions, there might be negligible background in the image.

We optimized the hyper parameters for model training with a grid search, to find out the best combination of  learning rate, dropout fraction, activation functions and kernel sizes. While augmentation is used in sparsely labelled or data-limited applications, it still serves as a useful technique for regularisation (avoiding overfitting). The combination of computationally expensive simulation with cheap training-time augmentation allows for a robust model. Flipping the inputs across different dimensions allows further robustness in deployability in case a user feeds the model an array with different dimension ordering. Finally, Histogram normalisation normalises brightness values which are usually skewed to the lower end to avoid photobleaching. A set of simulated and augmented images can be seen in Fig.~\ref{sFigColloidsAug} in the supplementary appendix.

\begin{figure*}
\includegraphics[width=\textwidth]{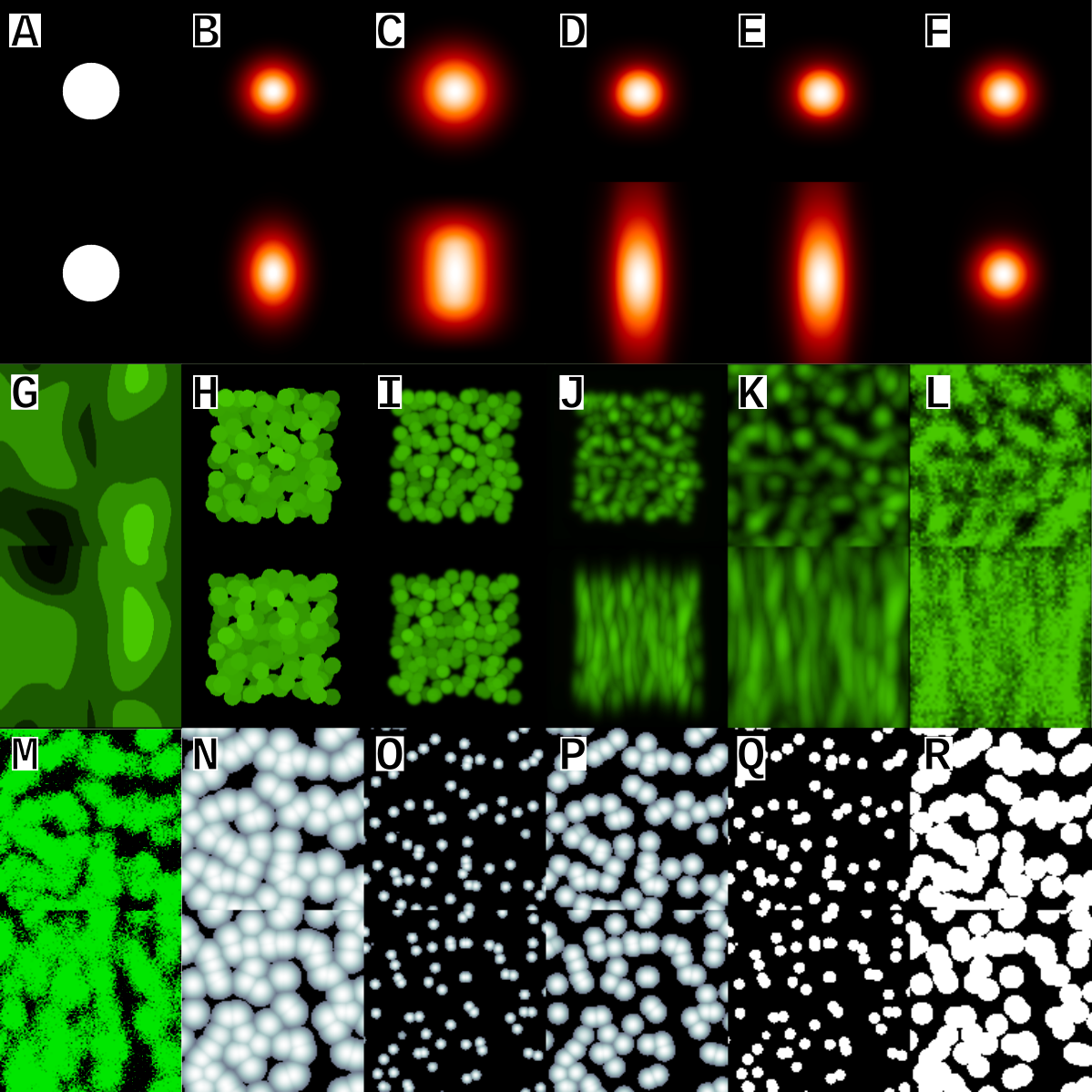}
\caption[Simulations.]{The process used in colloidoscope.
\textcolor{red}{Red}: PSFs. \textcolor{green}{Green}: Colloids,  \textcolor{gray}{White}: Training labels.
(a-f) Approximations of Confocal and STED PSFs.
(a) Simulated particle of diameter = 10 pixels. 
(b) The particle A) convolved with a Gaussian kernel of $(9,5,5)$ (z,x,y).
(c) A simulated particle of size 200 $\mathrm{nm}$ using the GBL method.
(d) A simulated particle of size 200 $\mathrm{nm}$ using a Huygens STED XY PSF.
(e) A simulated particle of size 200 $\mathrm{nm}$ using a Huygens STED 50XY 50Z PSF.
(f) A simulated particle of size 200 $\mathrm{nm}$ using a Huygens STED Z PSF.
(g-l) Colloid simulation steps.
(g) Background.
(h) Drawing spheres.
(i) Aliasing.
(j) Convolve PSF.
(k) Crop.
(l) Noise.
(m-r) Gaussian heatmap regression and semantic segmentation for particle detection.
(m) A simulated volume of colloids.
(n) Gaussian heatmap label of (A) with varying radius $r=\sigma$.
(o) Gaussian heatmap with constant radius $r=4$.
(p) Gaussian heatmap with constant radius $r=7$.
(q) Semantic labels of with varying radius $r=\sigma$.
(r) Semantic labels with constant radius $r=4$.}
\label{figSim}
\end{figure*}

\subsection{Simulation of the training dataset}
\label{sectionSimulation}

Due to the ground truth deficiency of this problem, the \unet must be trained purely on simulations. Each image can be different due to the system being imaged, the microscope, lens, or imaging parameters. The overall quality of an image is therefore an elusive measure. Brightness, noise, and contrast, combine to determine the quality of the image.  These are factors of the fluorescent dye, laser power, photobleaching, as well as detector sensitivity.

To begin, we generate particle positions for amorphous systems since colloidal crystals would change the optical properties of the system. Therefore, we focus on amorphous colloidal systems by using a hard-sphere Monte-Carlo algorithm~ ~\cite{allen,frenkel}. To begin many random positions are generated at a low volume fraction. Then the system is
``crushed'', where the volume is slowly decreased, with the particles being randomly moved until no overlap is measured. This process is then repeated until the desired volume fraction is reached. These coordinate sets were generated using Hoomdblue~\cite{anderson2016, anderson2020}.

After the positions are generated at different volume fractions, to create the simulation the experimental data parameters must first be measured.
The signal to noise ratio SNR is a useful measure for assessing signal quality. In signal analysis it is usually defined as the mean divided by the standard deviation. Since $\sigma$ refers to the particle diameter we define the \emph{sample} mean brightness as $\bar{x}$ with a standard deviation of $s$:

\begin{equation}
\mathrm{SNR} = \frac{\bar{x}}{s}
\label{eqSNR1}
\end{equation}

While in signal processing this is a pure measure given the signal is constant, in images this is a more qualitative measure. Instead define $f_{\bar{x}}$ and $f_s$ which are the \emph{foreground} mean brightness and noise for the images.

\begin{equation}
\mathrm{SNR} = \frac{f_{\bar{x}}}{f_{s}}
\label{eqSNR2}
\end{equation}

A more pertinent measure for object detection is the contrast to noise ratio (CNR). The CNR is commonly used in biomedical imaging to analyse the quality of different imaging modalities such as CT, MRI, PET, or ultrasound. Through various simulated datasets we found that the CNR is a salient measure for image quality, specifically for confocal imaging for colloids where tuning the laser illumination is important for photobleaching. This not only includes the mean of the foreground brightness and foreground noise, but also the background mean brightness $b_{\bar{x}}$. All the particles in the same system are not the same brightness, with the most challenging detections being particles that are extremely close to each other or touching. The separation of these two particles will depend on the noise, but more so on the contrast between the foreground and background.

We define the CNR as:

\begin{equation}
\mathrm{CNR} = \frac{| f_{\bar{x}} - b_{\bar{x}} |}{b_s}.
\label{eqCNR}
\end{equation}

With the CNR in hand, we proceed to generate simulated images from the coordinates obtained following the method above. We begin by convolving the `fluorescent cores' of the simulation with the point spread function to mimic the optics of the microscope. The PSF depends on many factors, including the excitation and emission wavelength, refractive-index of the particles and solution, numerical-aperture, magnification of the lens, as well as pinhole radius and shape. In practice, however, these factors have minimal effect on the size of the PSF. The ability to resolve particle coordinates ultimately rests on the size of the particle, but the PSF itself does not depend on particle size. The final blur is a function of the size of the PSF when compared to the size of the particle. The PSF of each lens system is of a constant size, but to investigate the effect of apparent particle radius in pixels, the simulated PSFs are resized by calculating the target \empty{pixels per nm}.

A crude method to approximate a PSF is using a Gaussian blur [Fig.~\ref{figSim}(b)], where the filter variance is larger in the $z$ dimension ~\cite{vanblaaderen1995}. An improvement is using a least-squares Gaussian approximation method [Fig~\ref{figSim}(c)]~\cite{zhang2007}. The same settings are used for all PSFs since parameters such as excitation wavelength contribute minimally to the PSF shape. An excitation wavelength of 488 nm was used. However, we are unaware of any methods for a least-squares approximation of a STED PSF.
Instead the PSFs were sourced from Huygens Professional version 22.10 (Scientific Volume Imaging, The Netherlands, http://svi.nl). Crucially for this project, Huygens software provides STED PSFs depending on the power of the $xy$ (doughnut) and $z$ depletion lasers.

Once the SNR and CNR are characterised, the positions are generated, and the PSF is ready, the simulated images
can be generated. This section will describe how the positions are used to draw a simulated image 
of size 64 pixels cubed seen in Fig. \ref{figSim} G-L. This simulation is simple but allows for easily parameterised datapoints. To generate such a simulated image, we proceed as follows:

\begin{itemize}
    \item We begin with an image twice as large as the target then later coarsen the image by a factor of two for aliasing. Padding is also added to allow the Gaussian blur and point spread function to be convolved in the image and to not lose the signal in the edges.
    \item Perlin noise~\cite{perlin1985} is used with 4 octaves to generate the background noise, the brightness of which is derived from the CNR measurements.~\footnote{Perlin noise corresponds to four greyscale values which are used for the image (Fig.~\ref{figSim} G)~\cite{perlin1985}.} 
    \item The positions and diameters are used to draw the particles in the image. 
    \item The image is zoomed in or resized for aliasing. (zooming is more computationally efficient than adding artificial aliasing). 
    \item The PSF is convolved with the simulated image. 
    \item The image padding is cropped and, 
    \item Finally noise is added.
\end{itemize}

The parameters of the training dataset are sampled from the distributions described in the supplementary appendix (Table~\ref{sTableTrainData}).

To isolate the effect of each simulation parameter, a randomly generated parameter set akin to the training dataset would be insufficient, since there would be too much noise in the data to isolate the effect of each parameter. A test dataset was created that isolates each parameter one-at-a-time. We choose to fix the other parameters at a level that remains at a challenging level for the \unet but not sufficiently to limit the performance. These values were chosen to represent the periods with the most dramatic fall in precision/recall of TP against each parameter ((Table~\ref{sTableTrainData}).).

\begin{figure*}
\centering
\includegraphics[width=180 mm]{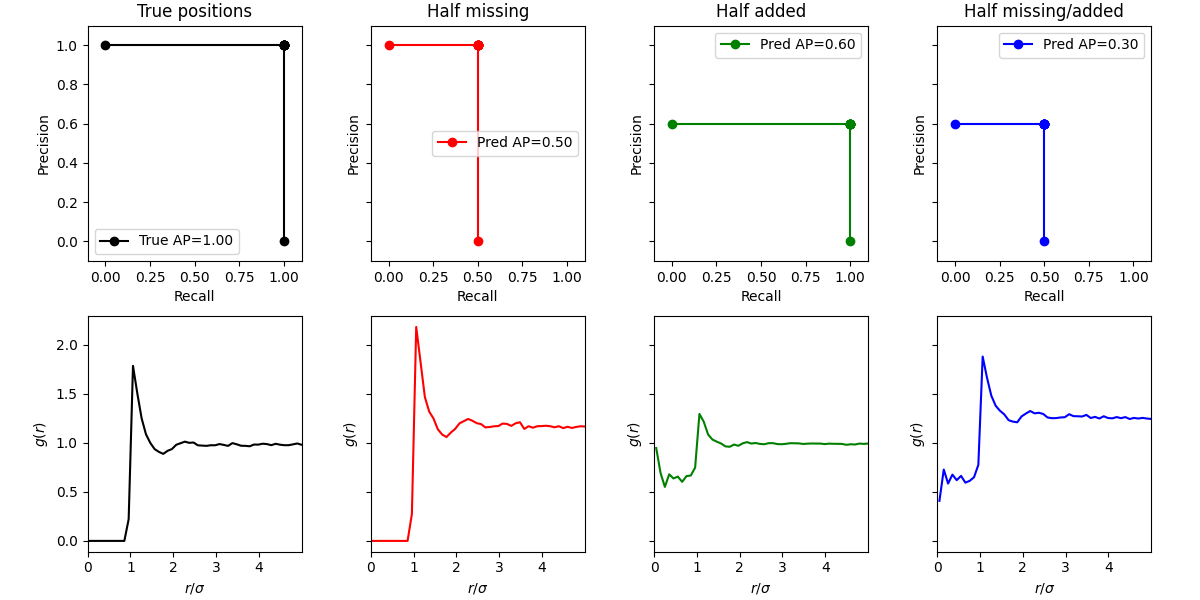}
\caption[Metrics]{Metrics: Average Precision (AP) and $g(r)$ of a simulated distribution. Removing half the particles (False negatives) results in a recall of $0.5$, this adds noise to the $g(r)$. Adding half of the particles as ghost particles (false Positives) results in a lower precision, and a signal in the $g(r)$ below $1$. Adding false positives and false negatives is also shown on the right.}
\label{figMetricsSim}
\end{figure*}

\subsection{Label simulation}
\label{sectionLabel}

Here the goal is to have high accuracy \emph{sub-pixel} resolution coordinates of all the particles in the image. Generating the labels for the simulation will depend on the manner we wish to use our model whether it be a semantic map (binary) or a continuous heatmap. Tested are approaches by many heuristics ~\cite{newby2018, midtvedt2020, midtvedt2022}.

To visualise examples, we simulate a random image and show multiple methods of drawing training labels in Fig. \ref{figSim} M-R. 
The label can either by a smooth Gaussian heatmap or it can be a binary mask. The radius of the label can equal the radius of the particle (Fig. \ref{figSim} N), a varying radius can suit object detection problems for polydispersed suspensions. On the other hand, the label can be a smaller fixed radius for every particle - akin to a bullseye (Fig. \ref{figSim} O and P). 
In Fig. \ref{figSim} Q and R shown are the usual binary semantic segmentation labels, representing foreground and background.

However, semantic labels overlap when it comes to closely packed targets. Therefore, we generate a Gaussian around each particle coordinate in the model. This Gaussian Heatmap Regression approach is taken to aid in dense detection. Furthermore, a smooth Gaussian heatmap permits the use of an $L_1$ loss function for the model. Regression loss functions can be more robust to data imbalance such as in very high or very low volume fractions when compared to binary mask pixel-wise classification.

\subsection{Model post processing}
\label{sectionostProcessing}

The direct output of our model is a heatmap which encodes the information of particle coordinates. We need to apply extra processing to convert such output to particle coordinates. For this conversion, we tested two approaches. Firstly we used the Crocker \& Grier's algorithm implemented in TrackPy (TP) to process the model output. We also used a Laplacian of Gaussian (LOG) algorithm to process the model output. See Figure \ref{figSchematic} for the different approaches taken.

\subsection{Accuracy metrics}
\label{sectionAccuracy}

As discussed previously by ~\citet{nibali2018} it is not possible to have a differentiable loss function derived from the Euclidean distances between the model predictions and the true locations since the number of particles can not be determined \emph{a-priori}.  Therefore, it is important to have intuitive and interpretable accuracy metrics that can offer an insight into model performance.

As noted above, to validate the model on simulated data where the ground truth is known, we use precision and recall. Precision describes the proportion of detections that are true. While recall measures the proportion of all true particles detected.

A common metric for object detection in machine learning is Average Precision (AP). However, here the important prediction is particle \emph{positions} rather than the bounding box, since particle size is already known even if polydispersed. AP uses intersection over union (IOU) - the IOU takes the intersection of the prediction with the ground truth divided by the union of both. Instead of IOU, we simply use the distance between the prediction and ground truth normalised by the particle diameter in pixels. AP is usually referred to as $AP_\tau$ where $\tau$ represents the percentage of the diameter as which a prediction is regarded as a match or mismatch. We therefore compute AP as follows:

\vspace{10pt}
Distance Matrix
\begin{equation}
D_{ij} = |\mathbf{p}_i - \mathbf{\hat{p}}_j|
\end{equation}

Mismatch Matrix
\begin{equation}
M_{ij} = \begin{cases}
1 & \text{if } D_{ij} > \tau \\
0 & \text{if } D_{ij} \leq \tau
\end{cases}
\end{equation}

Precision
\begin{equation}
\text{Precision} = \frac{1}{\hat{N}} \sum_{j=1}^{\hat{N}} \left( 1 - \prod_{i=1}^{N} M_{ij} \right)
\end{equation}

Recall
\begin{equation}
\text{Recall} = \frac{1}{N} \sum_{i=1}^{N} \left( 1 - \prod_{j=1}^{\hat{N}} M_{ij} \right)
\end{equation}

AP can be easily measured using the pairwise distances of the predictions and ground truths. The standard threshold is 50\% of the particle diameter ($AP_{50}$). When simply AP is referenced, this is the combined average precision of ten consecutive thresholds 0\% - 100\%.

If this is a ground truth deficient problem, how can accuracy be validated for experimental data? For this we leverage the simplicity of colloidal systems and the long history of structural measures applied to them~\cite{hansen1976,royall2023}. The Radial Distribution Function (RDF), $g(r)$, is uniquely determined for an isotropic fluid with spherically symmetric pairwise interactions and can be predicted from theory with high accuracy in many cases~\cite{hansen1976}.

In experiments, the $g(r)$ provides great insight into the nature of the prediction and can make biases in detection clear. For instance, if nearby particles are labelled too close, this will show as a peak before $r / \sigma = 1$ meaning a decrease in precision (more false positives, see Fig. \ref{figMetricsSim}). If the $g(r)$ does not level off at unity for large $r$, this can hint at long lengthscale inconsistencies in the detection such as clustering or boundary effects. This can then be used to validate the 
detections of both simulated and experimental data.

In addition to the pair correlation function ($g(r)$), finer details can be probed with \emph{higher-order} structural measures, which probe the correlations of multiple particles. A suitable method here is the \emph{Topological Cluster Classifcation} (TCC)~\cite{Malins2013}. This identifies groups of particles whose bond topology (identified by a Voronoi decomposition) is identical to isolated clusters. When processing the experimental results where the ground truth for particle locations are not available, TCC acts as a more sensitive measure for the PRS results compared to the RDF ~\cite{rattachai2017}.

\subsection{Colloidal Suspension Imaging}
\label{sectionSuspensionPreparation}

To validate our methods, it is important to evaluate the model on different \emph{experimental} image specimens. This is especially important since the training is only done on simulations. In this work five different  colloidal suspensions will be analysed. In particle--resolved studies of colloids, for refractive index--matched systems, early work used novel core-shell particles, where the fluorescent dye is located only at the centre of the particle ~\cite{vanblaaderen1995}. This acts as a means to improve the precision and recall of the detected particle coordinates.

Here, we consider a range of experimental samples of different densities, media, sizes, and fluorescent dyes. And to provide evidence of overfitting (or lack thereof) and support the discussion of generalisability in particle tracking applications. These suspensions are refractive index and density matched to neglect opaqueness and bouyancy effects. Described in 
Table~\ref{sTableRealCol} in the supplementary appendix are the composition, size, volume fraction, and polydispersity of the experimental data used to validate the model. A Leica SP8 3D STED II confocal microscope was used to image the specimens.

\section{Results}
\label{sectionResults}

In this results section, we evaluate the ability of the \unet to detect the coordinates of colloids in confocal microscopy images. We highlight the enhanced capability to distinguish colloids in \numcolloidschallenging challenging imaging scenarios when the model is used. We then showcase the model's performance on a relatively less challenging image. This marks a significant advancement in automating the analysis of confocal microscopy data, demonstrating the model's effectiveness in real-world applications.

\begin{figure*}[!ht]
\includegraphics[width=\textwidth]{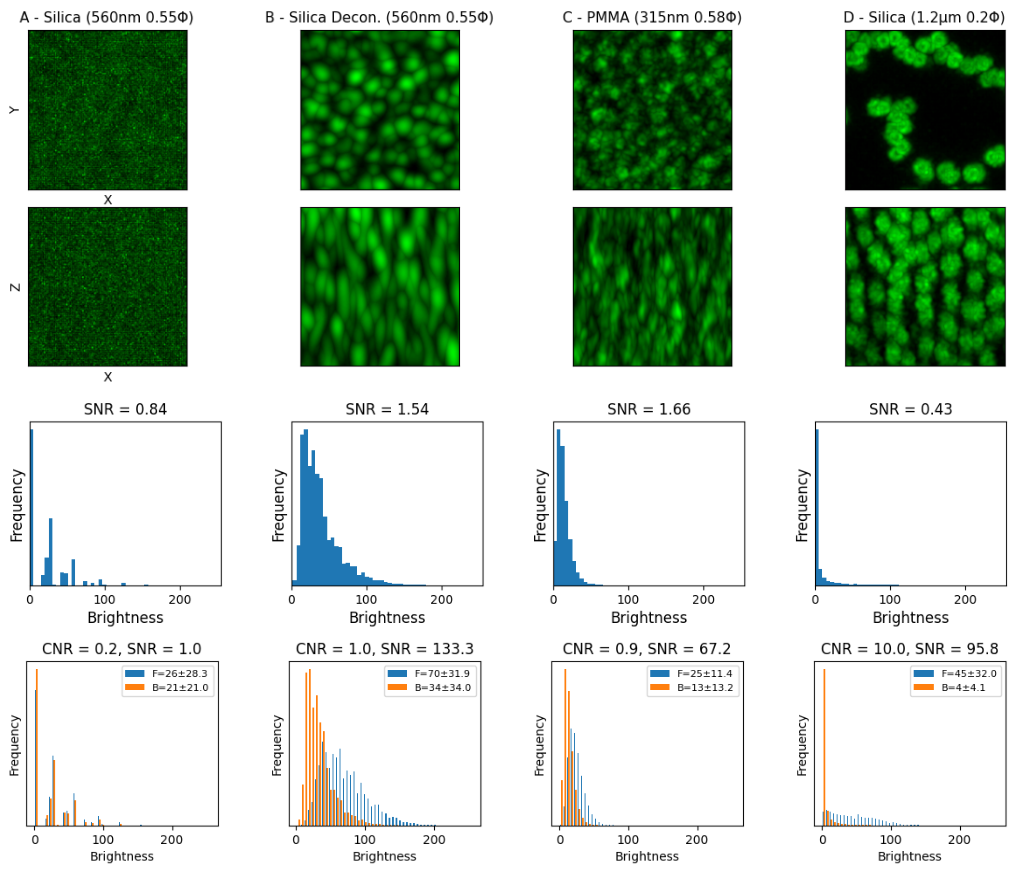}
\caption[Signal and contrast to noise ratio of experimental data.]
{Signal and contrast to noise ratio of experimental data.  The first and second row indicate $xy$ and $xz$ views of the colloid images. The raw brightness distributions and SNR ($\bar{x} / s$) of each image are shown in the third row. Finally the experimental images are separated into foreground (F) and background (B), whose brightness distributions are plotted independantly in the bottom row. The legends indicate the mean +/- standard deviation.}
\label{figExpSnrCnr}
\end{figure*}

\subsection{Experimental data measurements}

To tune the simulation it is important to obtain a benchmark from the experimental data. This is particularly true for the SNR and CNR. To begin the brightness distributions of each of the \numcolloidschallenging data points are plotted (See Fig.~\ref{figExpSnrCnr}). A crude measurement of signal to noise ratio $\bar{x}/s$ shows values between an SNR of 0.4 and 2. However, the distribution contains the brightness measurement for both background and foreground. This skews the distribution towards lower values and can be seen when there is particular imbalance in the data such the sheet-like structures in Fig.~\ref{figExpSnrCnr} D.

Labelling the particles and background separately allows the comparison of their probability distributions (See the bottom row of Fig. \ref{figExpSnrCnr}). Since the values are being measured experimentally, false negatives are unavoidable, skewing the distributions closer together. The PSF is anisotropic so drawing perfect spheres of radius $r$ would not capture the correct segmentation mask. Due to these challenges we use the median rather than the mean which is a more robust estimator to outliers. One can find the CNR to be between 0.2 and 10. The laser illumination power is usually tuned so that particles are dim, far from the maximum of 256 or 65536 for 8 or 16 bit respectively. While this might require increasing the gain, reducing illumination has 3 advantages. It avoids photobleaching, eliminates overexposing past the sensor maximum, and improves signal to noise.

\begin{figure*}[!ht]
\includegraphics[width=\textwidth]{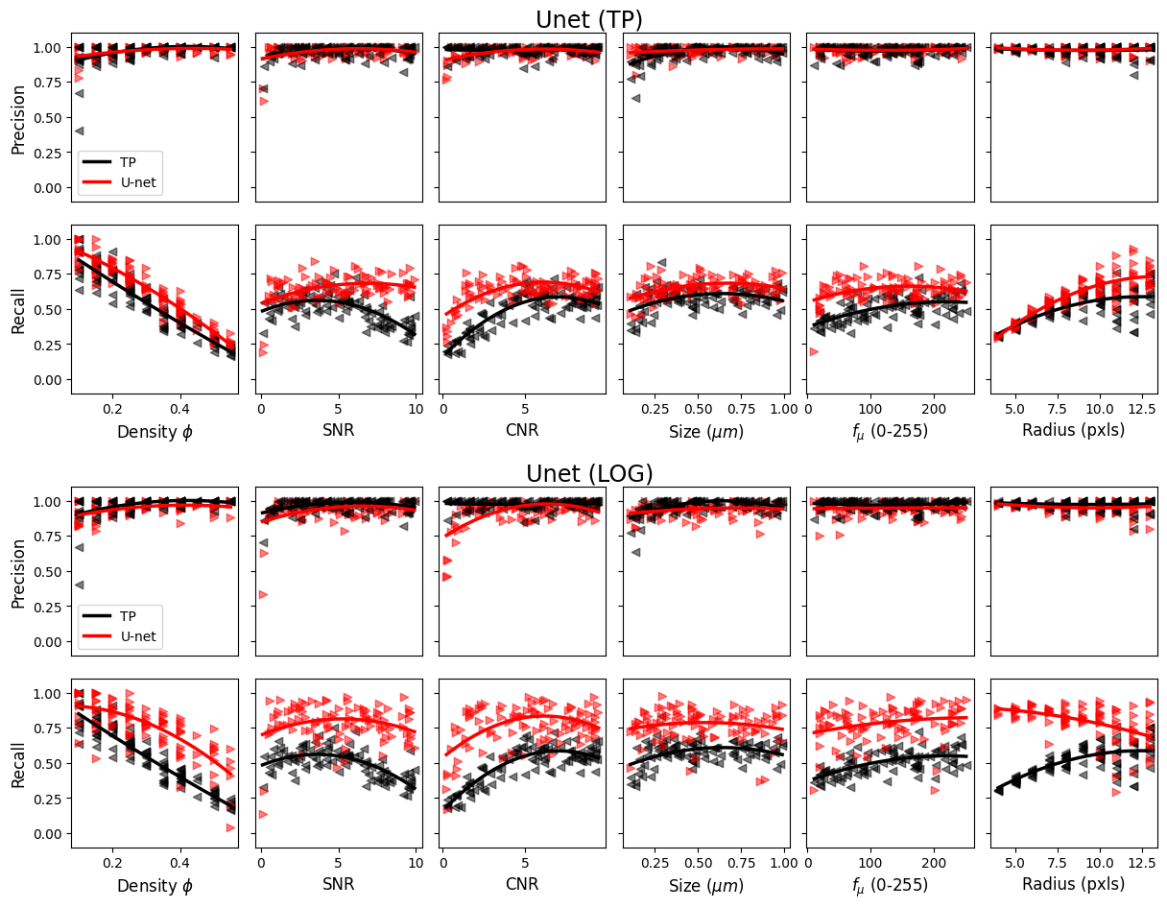}
\caption[\unet]{\unet precision and recall ($AP_{50}$) against simulation parameters. The top and bottom figures show the results when using Trackpy (TP) and Laplacian of Gaussian (LOG) respectively to extract particle positions from the model labels.}
\label{figPR}
\end{figure*}

\subsection{Evaluation - \unet}

Over 20 datasets were simulated from the contrast and noise measurements to result in the final steps and parameter distributions described in the methods. The \unet
has been trained over 1000 times accross multiple grid searches. A 3D residual \unet with a block size of 2 (equivilant to a Resnet18 layer encoder ~\cite{drozdzal2016}) was trained with a learning rate of 0.002 for 5 epochs. A kernel size of 7 improved accuracy particularly on large particles. Finally the activation was set to a SWISH function~\cite{ramachandran2017}, and instance normalisation~\cite{ulyanov2016} served as the activation function. Finally, the dropout was set to 0.2.

Using a heatmap approach, firstly the model is tested as a ``denoiser'' for TP, wherein the model labels are fed to TP for post-processing as shown in Fig. \ref{figPR}. Over a parameter sweep TP maintains almost perfect precision. However, TP suffers from low recall usually only detecting 50\% of the particles due to its assumptions and refinement steps. Particularly it fails in low brightness and dense systems where particle separability suffers.

The \unet improves the recall from TP across all simulation parameters while maintaining almost perfect precision. The volume fraction ($\phi$) is the main objective and the \unet improves recall at 0.55 $\phi$ from 20\% to 30\% in simulations. The biggest improvements are in contrast and brightness, this is due to TP assuming the particles are in the brightest 30\% of the image. While this parameter can be tuned manually and TP can usually perform well at low brightness, the \unet provides a simpler interface during inference and doesn't require manual tuning. Interestingly, both \unet and TP suffer on particles smaller than 10 pixels in radius.

\begin{figure*}[!ht]
\includegraphics[width=\textwidth]{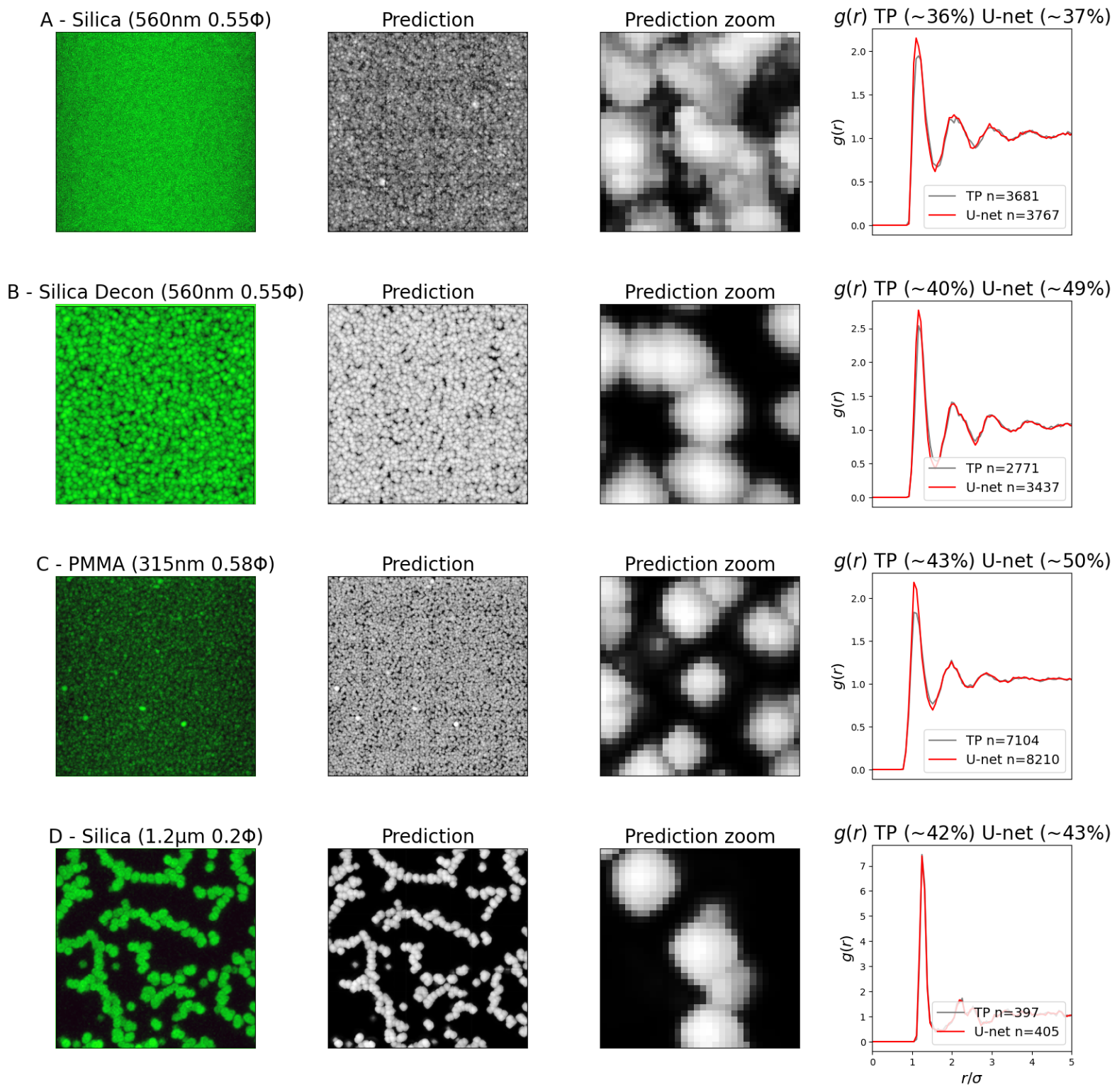}
\caption[\unet and TP on real data.]{Experimental predictions Using TP as a postprocessing method for a \unet.
(a-b) Experimental images of colloids. (b) Deconvolved (Decon.) volume of (a).
The approximated percentage detected is derived from the experimentally prepared and predicted volume fractions.
}
\label{figExperimental}
\end{figure*}

The limited recall of TP motivates the use of a different postprocessing method. We used a Laplacian of Gaussian (LOG) blob detection algorithm which shows a tradeoff of slightly lower precision for a pronounced higher recall accross simulated image parameters (See the bottom row of Fig.~\ref{figPR}). Most notably LOG rescues the low recall on particles smaller than 10 pixels. LOG is also less consistent than TP, with a larger spread of recall. This hints at a lower quality of segmentation labels. The LOG algorithm does not contain any refinement steps, meaning the detections are more reliant on the quality of the model's heatmap labels.

\subsection{Evaluation on experimental 
data}
\label{sectionExperimental}

How do these improvements carry forwards with experimental data? While the actual precision and recall can not be measured due to the ground truth deficiency, the the total number of particles detected and $g(r)$ can be used to analyse the quality of the prediction. Figure \ref{figExperimental} shows that there are more particles detected by the \unet, suggesting that the recall measured using Trackpy is again improved in every example of experimental data. More importantly we find the $g(r)$ shows no signs of false-positive detections, since there is no signal in the radial distribution lower than the peak $r/\sigma < 1$.

The overall results, increased number of detected particles and the physically meaningful radial distribution functions, indicate that the Image $\rightarrow$ \unet $\rightarrow$ TP pipeline improved the PRS result across different experimental conditions.

In experimental data postprocessing with Laplacian of Gaussian shows improved precision, but struggles on images of polydisperse systems (Fig.~\ref{sFigUnetLOGReal}C \& D) and dilute suspensions (Fig.~\ref{sFigUnetLOGReal} E). The radial distribution functions obtained from experimental data have higher first peaks in Fig.~\ref{sFigUnetLOGReal} A - C, which suggests the model is detecting more of the particles closely packed together, furthermore, the $g(r)$ stabilises around 1, indicating there are no long range inhomogeneities in particle detection.

\subsection{Evaluation: \attunet with Laplacian of Gaussian}

To attempt to translate the benefits of Laplacian of Gaussian on simulations to experimental data, a different model is chosen to see whether increased heatmap quality can boost recall and precision. \attunet claims to improve segmentation quality by guiding the models output towards important pixels. It conserves more spatial information from the encoder to the decoder, and highlights the dense features important for detection. For colloid tracking an \attunet with LOG postprocessing shows very similar results on simulated data to a \unet - see Fig.~\ref{sFigAttUNetLogReal}. \attunet does not show an improved recall on experimental data over a normal \unet. However, there is higher contrast in the predicted heatmaps which could make it easier for the post processing method. Overall, we find \attunet improves the first peaks of the experimental pair correlation functions (Supplementary Fig.~\ref{sFigAttUNetLogReal} A-C).

\begin{figure*}
\centering
\includegraphics[width=0.7\textwidth]{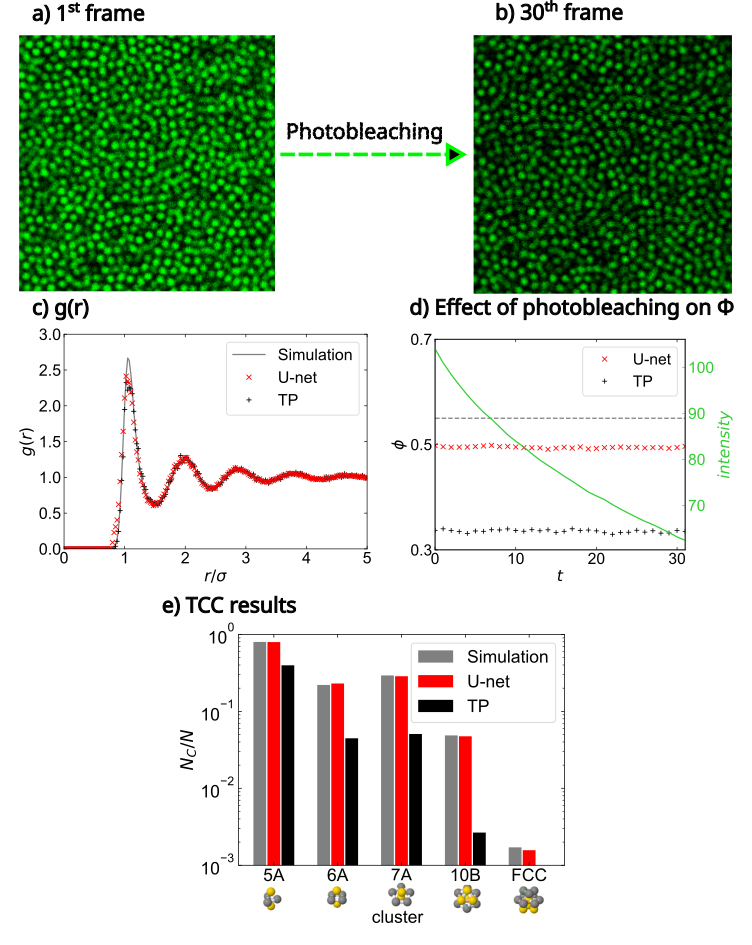}
\caption[Photo-bleaching and TCC results] {The \unet applied to larger colloids: 
Photo-bleaching and topological cluster classification results.
(a-b) Show the effect of consecutively imaging a volume of colloids for 30 frames.
(c) Quantified intensity (brightness) of the 30 images, with the predicted volume fraction ($g(r)$) of the \unet and Trackpy (TP).
(d) A volume of colloids can be simulated at the volume fraction ($\phi = 0.55$) of the prepared samples. This simulation can then be used to compare with the experimental $g(r)$.
(e) TCC (Topological Cluster Classification) results on simulated and experimental data quantified with a \unet and Trackpy.
}
\label{figLars}
\end{figure*}

\subsection{Evaluation using larger colloids: Photobleaching and higher-order structural measures}

Finally, after fine tuning and testing the model on the smaller particles, we choose a relatively simpler image of 2 micron particles at $\phi = 0.55$ with a polydispersity of 4\%. First, we analysed the effect of photobleaching, i.e. the loss of signal to noise ratio and decrease in average brightness of the image over the course of a measurement of 32 frames (see Fig.~\ref{figLars}A and B).

The \unet maintains higher recall without impairing precision as shown from the $g(r)$ in Fig. \ref{figLars} C and the comparison of the calculated volume fraction by counting particles in Fig. \ref{figLars} D. The high precision coordinates produced by Colloidoscope result in a higher first peak that perfectly matches the shape of the simulation $g(r)$. With the intensity of the image falling from 150 to 50 (in 8 bit brightness) Trackpy predicts a constant volume fraction of $\phi = 0.35$, while colloidoscope maintains a prediction of $\phi = 0.5$, much closer to the ground truth of $\phi = 0.55$.

So far, we have considered pair correlations of particles through the radial distribution function $g(r)$. However, higher–order structural correlations provide a more detailed probe of the quality of particle tracking when benchmarked against simulation data. Correctly classifying these structures requires high accuracy detection. For example, in the case of the fcc cluster, a failure to identify one particle out of 13 will result in the cluster not being detected.

We calculated the averaged population of different cluster types identified from coordinates produced by Colloidoscope and Trackpy and compared them to hard sphere simulation data. Here we consider five clusters each consisting of $m$ particles relevant to the system in question. These are the triangular bipyramid ($m = 5$), octahedron ($m = 6$), 7-membered pentagonal bipyramid, defective icosahedron ($m = 10$) and face-centred cubic crystal ($m = 13$). We see (Fig. \ref{figLars} E)) that Colloidoscope matches the cluster population of the simulation, maintaining a closer match than TP. This shows great promise for this tool in delivering higher quality coordinates than otherwise possible.

\section{Discussion}

The \unet consistently outperforms heuristics on both simulated and experimental data in different configurations. Particle resolved studies
of confocal microscopy data for applications of soft matter physics requires the highest precision achievable. This work shows that deep learning approaches can achieve the robustness required for this analysis. Increases in precision and recall using the algorithm described directly lead to improvements in the $g(r)$ of experimental data.

The findings from simulated data generally translated to experimental data, images usually contain a bad mix of all image parameters (e.g. high density, high noise, and low contrast). This shows the challenging nature of designing simulation datasets. The simulated training dataset included volume fractions up to $\phi=0.55$.  Nevertheless the model still generalised well to experimental data up to 0.58 $\phi$ (See Fig. \ref{figExperimental} C).

Image quality is an elusive measure and combines all the parameters discussed here. While the diffraction limit has been broken for single fluorophores, which may be detected to an accuracy of 20 nm in STORM imaging ~\cite{nehme2018}. However, tracking in dense suspensions of colloids remains more challenging. Deconvolution can introduce artefacts which are challenging to simulate, with the model often failing to improve TP results on the deconvolved experimental data (See Fig. \ref{figExperimental} A vs B) where, for example, colloidoscope detects more particles (3767) from a raw image than Trackpy on a deconvolved image (2771).

Trackpy struggles with polydispersity since it needs the apparent diameter as an input parameter, since its underlying algorithm assumes all particles to be detected have the same size.
LOG can detect particles within a range which shows its superior performance on large and small simulated particles, as well as polydispersed volumes (See the right most plots in Fig. \ref{figPR}). The processing methods TP and LOG can give further insight into the performance of the model heatmap. In this case the \unet improved the performance of both postprocessing methods. For dense inference, the \attunet further matched the high precision of TP, with the benefits of LOG recall. While attention acts to increase model performance in situations with data imbalance (dilute suspensions), it also boosts performance on polydispersed data (Fig.~\ref{sFigAttUNetLogReal}).

This model is deployed and available to use pretrained \url{https://github.com/wahabk/colloidoscope}. We choose a standard \unet over an \attunet since we find it gives better results on 
data taken from suspensions at high volume fraction. The only parameter that is required is the diameter of the particles. The users are given the option of using TP or LOG, with the choice dependant on whether high precision or high recall is required, often LOG aids in detecting polydispersed colloids. This shows how Deep Learning can also be more accessible in deployment than previous methods. While DeepTrack ~\cite{midtvedt2022} can detect many shapes of colloids (such as crescents), it requires the user to simulate their own data and train their own model, significantly hampering user-friendliness and adoption. This model is deployed pretrained and confidently generalises to all colloidal spheres tested.

\subsection*{Limitations}

Due to the inability to determine the number of particles \emph{a priori} this model was not trained end-to-end (Fig. \ref{figSchematic}). Furthermore, another limitation of this approach is tracking by detection - i.e. every frame is analysed separately. Integrating the information across frames can improve model performance if a particle is out of view for only one frame ~\cite{gehrig2023}. This would also aid in training, since the model is also learning the temporal information and could lead to a better inductive bias. A recurrent approach would also aid in deployment, removing the requirement for further postprocessing steps - mainly trajectory linking. However, the practical training of a recurrent network requires back propagation through time (BPTT) which,  prohibitively, requires much more training compute budget.

Further network architectures are being developed to counter the difficulty in instance segmentation using embedding. The issue of determining the number of particles before prediction can be phrased in a second way. It is challenging to know which segmentations (foreground)  belongs to which particle. This is commonly named the data association problem, and is countered using output embedding. DSNT (Differentiable Spatial to Numerical Transform~\cite{nibali2018}) embeds the output by predetermining which particle class goes in which output channel during training. Further approaches counter this by semantically labelling the input, then discriminating different instances using clustering approaches~\cite{brabandere2017,neven2019}.

While an \attunet was used, this model does not benefit from full self attention such as traditional transformers. Transformers were created to address limitations in maintaining context in Recurrent Neural Networks. Furthermore, transformer architectures benefit from positional encoding of patches, instead of analysing patches without context of surrounding ones. Embedding of the input and output can act to alleviate many of the limitations encountered here~\cite{hu2021}. However, transformers still require large datasets ranging from hundreds of thousands to hundreds of millions which is beyond our compute budget ~\cite{dosovitskiy2021}.

\section{Conclusion}

In this study, we proposed a neural network based particle resolved study pipeline, Colloidoscope. We compared colloidoscope with existing method TrackPy (TP) for their applicability in imaging dense colloidal suspensions.

\begin{itemize}
    \item Colloidoscope outperformed TP by maintaining high precision and improving recall on simulated and experimental data.
    \item Colloidoscope demonstrated the ability to improve predictions in high colloid volume fraction
    and low contrast, thus expanding the scope of imaging and effectively cured the effect of photobleaching when images are collected over a long period of time.
    \item  Furthermore, colloidoscope offers a more user friendly interface with one single parameter, compared to TP. Colloidoscope easily generalized to multiple suspensions (both silica and PMMA colloids), and simply using raw images Colloidoscope outperforms Trackpy on deconvolved images (See Fig. \ref{figExperimental} A vs. B).
    \item Overall, the findings suggest that Colloidoscope is a promising alternative to TP for imaging Colloidal suspensions with enhanced performance and user-friendliness.
\end{itemize}

\section*{Author Contributions}
This project was conceived by AK, YY and CPR. 
AK implemented the \unet and analysed the data. 
YY devised the simulations for the training of the \unet.
LK, LO, JEH, and AMA contributed data.
All authors wrote the manuscript.

\section*{Conflicts of interest}
There are no conflicts to declare.

\section*{Acknowledgements}
The authors would like to thank Giulio Biroli for very many stimulating discussions.
AK acknowledges the SWDTP for funding.
YY acknowledges funding support from the China Scholarship Council. 
LK and CPR acknowledges the Agence Nationale de Recherche for grant DiViNew 
CPR and AMA were supported by EPSRC grant EP/T031077/1.



\renewcommand\thetable{S\arabic{table}}
\renewcommand\thefigure{S\arabic{figure}} 
\setcounter{figure}{0}

\begin{table*}[h]
\begin{tabular}{llllll}
  & Particle & Size         & \begin{tabular}[c]{@{}l@{}}Denisty\\ $\phi$\end{tabular} & Description & Acknowledgement \\ \hline
A & Silica   & 560 nm     & 0.55   &  \begin{tabular}[c]{@{}l@{}}Silica in DMSO \\ Monodispersed \end{tabular}                                                                                                                                           & James Hallet         \\ \hline
B & Silica   & 560 nm     & 0.55   &  \begin{tabular}[c]{@{}l@{}}Silica in DMSO \\ (deconvolved) \end{tabular}                                                                                                                                            & James Hallet        \\ \hline
C & PMMA     & 315 nm     & 0.58   & \begin{tabular}[c]{@{}l@{}}Polydispersed PMMA \\ in dodecane \\ (Deconvolved)~\cite{hallett2018} \end{tabular} & Levke Ortlieb  \\ \hline
D & Silica   & 1.2  $\mu$m & 0.2    & \begin{tabular}[c]{@{}l@{}}Dipolar silica active \\ colloids with metallic \\ hemisphere - in ethanol\\  and DMSO~\cite{sakai2020} \end{tabular} & \begin{tabular}[c]{@{}l@{}}Katherine\\ Skipper\end{tabular}  \\ \hline
E & PMMA   & 2 $\mu$m & 0.55   &  \begin{tabular}[c]{@{}l@{}} PMMA in Cyclohexyl \\ bromide, decalin, and tetralin. \end{tabular}                            & Lars K\"{u}rten        \\ \hline
\end{tabular}
\caption{
Experimental colloids used in model evaluation.}
\label{sTableRealCol}
\end{table*}

\begin{table*}
\begin{tabular}{||cccccc||}
	\hline
	\begin{tabular}[c]{@{}l@{}}Radius \\(pixels)\end{tabular}   & \begin{tabular}[c]{@{}l@{}}Particle \\ Size \\  ($\mu m$)\end{tabular} & \begin{tabular}[c]{@{}l@{}}$f_{\mu}$ \\ (8bit)\end{tabular} & CNR           & SNR           & \begin{tabular}[c]{@{}l@{}}Density \\ ($\phi$) \end{tabular}    \\ \hline\hline
	(4,14) 		  & (0.01,1)      		 & (30,200)  		& (2,10) 		& (2,10) 		& (0.2,0.55) \\ \hline
\end{tabular}
\caption{Distributions of the simulated training data.}
\label{sTableTrainData}
\end{table*}

\begin{table*}
\centering
	\begin{tabular}{||c|cccccc||} 
		\hline
		\begin{tabular}[c]{@{}l@{}} $P_{fixed}$ / \\ $P_{analyse}$ \end{tabular}			& \begin{tabular}[c]{@{}l@{}} Rad. \\(pxls)\end{tabular} 			& \begin{tabular}[c]{@{}l@{}} Part. \\Size \\($\mu m$) \end{tabular} 			& $f_{\mu}$ (8bit) 				& CNR 					& SNR 				& \begin{tabular}[c]{@{}l@{}} Dens. \\ ($\phi$)\end{tabular} \\ [0.5ex] 
		\hline\hline
		\begin{tabular}[c]{@{}l@{}} Radius \\ (pixels) \end{tabular}							& (4,14) 				& 1 					& 255 						& 5 						& 5 					& 0.3 \\ 
		\hline
		\begin{tabular}[c]{@{}l@{}} Particle \\ size \\($\mu m$) \end{tabular}					& 10 					& (0.1,1)				& 255 						& 5 						& 5 					& 0.3 \\
		\hline
		$f_{\mu}$ (8bit) 														& 10 					& 1 					& 10,255					& 5 						& 5 					& 0.3 \\
		\hline
		CNR 																	& 10 					& 1 					& 255 						& (0.1,10)					& 5 					& 0.3 \\
		\hline
		SNR 																	& 10 					& 1 					& 255 						& 5 						& (0.1,10)				& 0.3 \\ [1ex] 
		\hline
		\begin{tabular}[c]{@{}l@{}} Density \\ ($\phi$) \end{tabular}							& 10 					& 1 					& 255 						& 5 						& 5 					& (0.1,0.55) \\ [1ex] 
		\hline
	\end{tabular}
\caption{One-at-a-time. Diagonal distribution of parameter sweeps. This allows the investigation of the effect of each parameter separately.}
\label{sTableValData}
\end{table*}

\begin{figure*}[h]
\includegraphics[width=\textwidth]{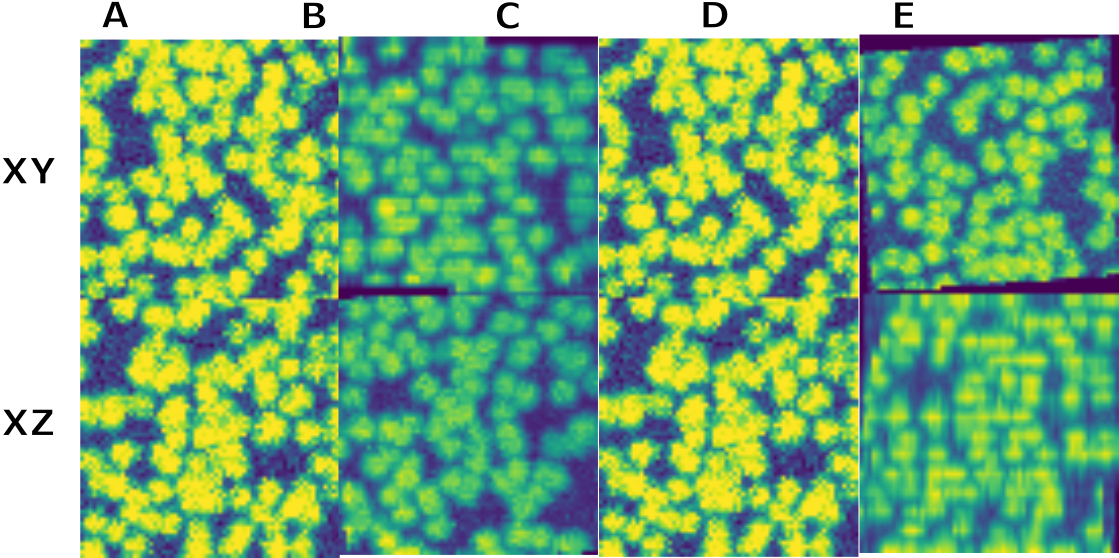}
\caption[Data augmentation.]{Examples of the data augmentation transformations used.
(a) Simulated volume of colloids.
(b-d) Augmentations of (a) using the described parameters below.
}
\label{sFigColloidsAug}
\end{figure*}

\begin{figure*}[]
\includegraphics[width=\textwidth]{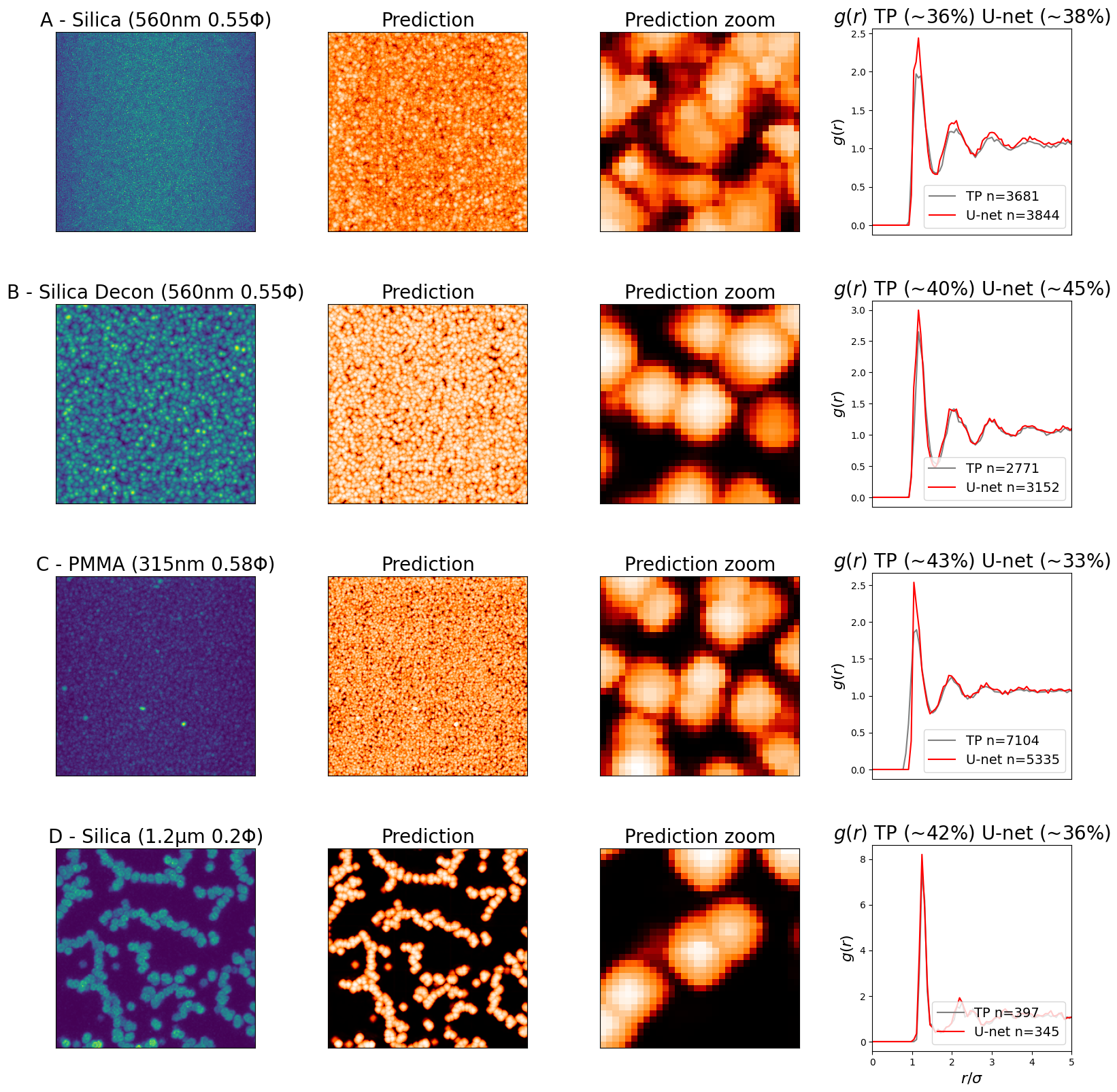}
\caption[U-net and LOG on real data.]{Real data predictions Using Laplacian of Gaussian (LOG) as a postprocessing method for a U-net.}
\label{sFigUNetLogReal}
\end{figure*}

\begin{figure*}[]
\includegraphics[width=\textwidth]{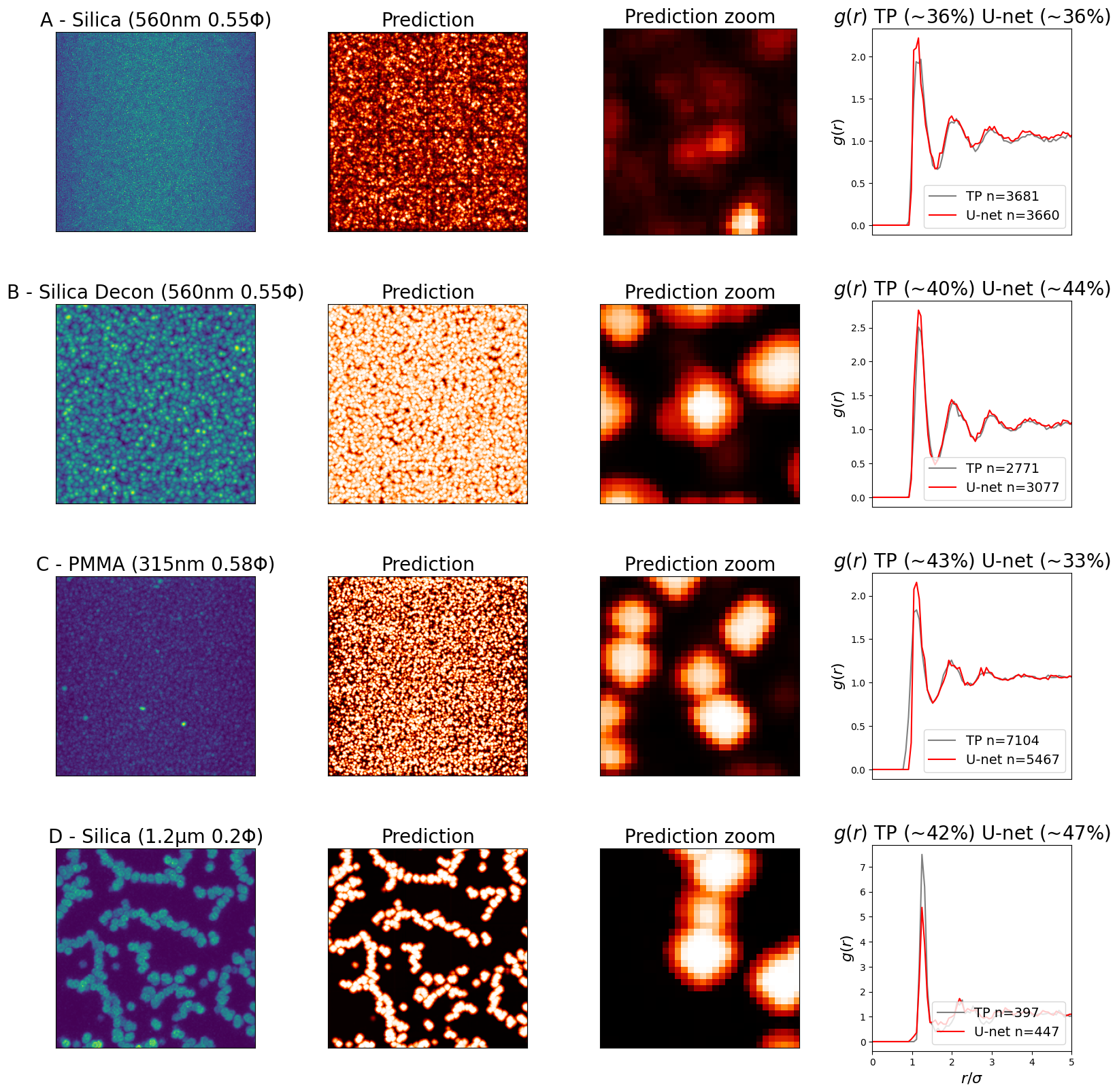}
\caption[Attention U-net and LOG on real data.]{Real data predictions Using Laplacian of Gaussian (LOG) as a postprocessing method for a attention U-net.}
\label{sFigAttUNetLogReal}
\end{figure*}

\end{document}